\documentclass[12pt]{article}
\usepackage{graphicx}

\def\Pisa{INFN Sezione di Pisa\\
Largo B. Pontecorvo 3, I-56127 Pisa, ITALY}

\def\Title#1{\begin{center} {\Large #1 } \end{center}}
\def\Author#1{\begin{center}{ \sc #1} \end{center}}
\def\Address#1{\begin{center}{ \it #1} \end{center}}

\newenvironment{Abstract}{\begin{quotation}  }{\end{quotation}}
\newenvironment{Presented}{\begin{quotation} \begin{center} 
             PRESENTED AT\end{center}\bigskip 
      \begin{center}\begin{large}}{\end{large}\end{center} \end{quotation}}
\def\Acknowledgements{\bigskip  \bigskip \begin{center} \begin{large}
             \bf ACKNOWLEDGEMENTS \end{large}\end{center}}

\begin{document}
\begin{titlepage}
\vfill
\Title{Charged Lepton Flavor Violation Experiments}
\vfill
\Author{Giovanni Signorelli}
\Address{\Pisa}
\vfill
\begin{Abstract}
The experimental status of charged lepton flavor violation searches is briefly reviewed, 
with particular emphasis on the three classical searches involving muon transisions: $\mu
\to e \gamma$, $\mu \to e$ conversion and $\mu \to 3e$. 
\end{Abstract}
\vfill
\begin{Presented}
2013 Flavor Physics and CP Violation (FPCP-2013),\\
Buzios, Rio de Janeiro, Brazil, May 19-24 2013
\end{Presented}
\vfill
\end{titlepage}
\def\thefootnote{\fnsymbol{footnote}}
\setcounter{footnote}{0}

\section{Introduction}
Charged lepton flavor transitions are forbidden in the Standard Model because of the vanishing
neutrino masses. The introduction of neutrino masses and mixing induces flavor transitions
radiatively, but at a negligible level. In fact the expected branching ratio for $\mu \to e
\gamma$ with the present knowledge on neutrino mixing parameters is $\approx 10^{-51}$.
The observation of charged lepton flavor transitions would represent a clear signal
of physics beyond the Standard Model, being virtually background free.

This situation is completely different from that in the quark sector where flavor transitions  
are already present through the Cabibbo-Kobayashi-Maskawa mixing matrix, and signals of 
new physics would appear as deviations from the predicted branching ratios, whose computation 
is not always free from theoretical difficulties.

In what follows I will review the current experimental status and near-future perspectives 
of the ``classical'' searches 
involving muon transitions, namely the $\mu \to e \gamma$ decay, the $\mu \to e$ conversion
and the $\mu \to 3e$.

\section{Charged Lepton Flavor Violation}
Charged Lepton Flavor Violation (CLFV) is related to new couplings and can be described in 
the effective operator language by dimension-5 or dimension-6 operators
\begin{equation}
\frac{1}{\Lambda} \bar \ell_i \sigma_{\mu\nu} \ell_j F^{\mu\nu} 
\qquad
\frac{1}{\Lambda^2} \bar \ell_i \gamma_\mu \ell_j 
\left( \bar q_k \gamma^\mu q_m + \bar \ell_k \gamma^\mu \ell_m \right)
\end{equation}
where $\Lambda$ is the effective scale, $\ell_i$ and $q_i$ are the leptons and quarks 
of different generations. The dimension-5 operator summarizes the couplings with the 
electro-magnetic field, and it is responsible of the $\mu \to e \gamma$ vertex present
also in tree-level $\mu \to e$ and $\mu \to 3e$ transitions, while 
the dimension-6 operators enter in the four-fermion vertices responsible for 
the two latter processes, but gives no contribution to the $\mu \to e \gamma$ decay.
 
The same effective operators describe also flavor-diagonal transitions (such as the muon 
and electron anomalous magnetic moment) and lepton flavor violating decays of taus
and flavored mesons.

It is in fact possible to correlate, in a model dependent way, many of these processes,
making it extremely interesting the investigation of different decays (see, 
{\em e.g.},~\cite{degouvea} and references therein). In Table~\ref{tab:theory} we show the main CLFV processes
with their naive probability scaling, their present limits on the year of the last measurement.

\begin{table}[t]
\begin{center}
\begin{tabular}{l|cccc}  
Process &  Relative probability & Present Limit & Experiment & Year \\ \hline

$\mu \to e \gamma$                &  1               &  $5.7 \times 10^{-13}$ & MEG        & 2012 \\
$\mu^- {\rm Ti} \to e^- {\rm Ti}$ & $Z \alpha / \pi$ &  $4.3 \times 10^{-12}$ & SINDRUM II & 2006 \\ 
$\mu^- {\rm Au} \to e^- {\rm Au}$ & $Z \alpha / \pi$ &  $7 \times 10^{-13}$   & SINDRUM II & 2006 \\ 
$\mu \to eee$                     & $\alpha / \pi$   &  $4.3 \times 10^{-12}$ & SINDRUM    & 1988 \\ 
$\tau \to \mu \gamma $  & $(m_\tau/m_\mu)^{2 \div 4} $ & $3.3 \times 10^{-8}$ & B-factories & 2011 \\
$\tau \to e \gamma $    & $(m_\tau/m_\mu)^{2 \div 4} $ & $4.5 \times 10^{-8}$ & B-factories & 2011 \\
\end{tabular}
\caption{Relative sensitivities and experimental limits of the main CLFV processes.}
\label{tab:theory}
\end{center}
\end{table}

The experimental effort can be divided in two categories: the ``exotic searches'', 
that is processes that if seen could indicate the existence of physics beyond the SM, and are
generally limited by the experiment side, and ``beyond SM physics'', {\em i.e.} processes
where new physics would appear as deviations from SM predictions, and are generally 
theory-limited. 
In some cases such processes can be searched for by multi purpose experiments (as in the case
of the B-factories) but sometimes dedicated experiments are mandatory, due to the extreme
specialization of the detector and to the performance requirements.
 
\section{The classical searches}
\begin{figure}[htb]
\centering
\includegraphics[width=0.8\columnwidth]{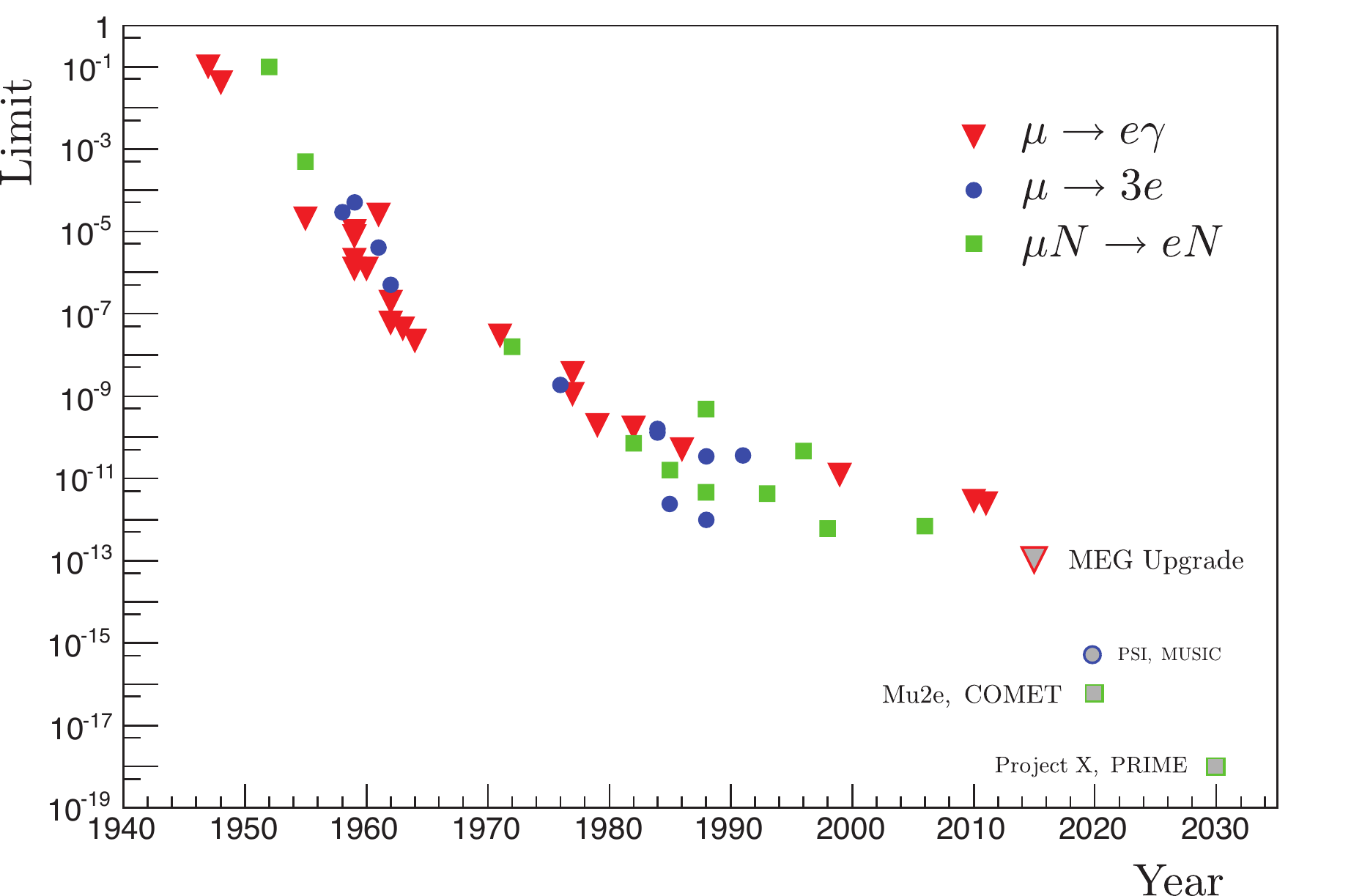}
\caption{The history of CLFV search in processes involving muons (after~\cite{Bernstein-Cooper}).}
\label{fig:history}
\end{figure}

In this paper I will concentrate on the three ``classical'' searches of CLFV decays involving muons,
which fall in the cathegory of dedicated experiments for exotic searches. They are $\mu \to e 
\gamma$, $\mu \to 3e$ and $\mu \to e$ conversion on nuclei.
In Figure~\ref{fig:history} we show the evolution of the limits set on this processes along the last 65
years, where we can see the three groups of experiments done with cosmic-ray muons (1940s) stopped 
pion beams (until mid-60s) and stopped muon beams (1970s onward). Each experiment proved to be an 
improvement over the previous one in either beam or detector technology.

\subsection{Kinematic and backgrounds}
The three processes involving muons share common characteristics, but each one shows a peculiarity that
makes it impossible to have a common experiment to search for all three simultaneously.

The $\mu \to e \gamma$ decay is a two body decay where the daughter particles are monoenergetic 
(52.8~MeV) and emitted simultaneously back-to-back in the muon rest frame. It is natural therefore to stop 
the muons in a thin target and for this reason a beam of positive muons is necessary, since negative
muons would undergo nuclear capture before decaying.
Two background processes can mimic a signal event: a muon radiative decay $\mu^+ \to e^+ \nu \bar \nu
\gamma$ in which the two neutino carry little energy and both positron and photon are close to their 
kinematic edge, and an accidental coincidence between a positron from a normal muon decay 
(``Michel positron'') and a high energy photon coming from a radiative decay, bremsstrahlung or positron
annihilation in flight. It is possible to show~\cite{kuno-okada} that for competitive resolutions the 
dominant background is the accidental one. For this reason a continuous muon beam is preferred, since
the probability of random coincidences from daughter particles coming from different muons is minimized.

In the $\mu {\cal N} \to e {\cal N}$ transition monoenergetic electrons are emitted against the spectator
nucleus, with an energy equal to the muon rest mass minus the muon binding energy. A single particle
is present in the final state and the main background events come from muon decay in orbit (DIO) or 
from interactions of particles present in the primary muon beam. A beam of negative muons is needed
to have the formation of muonic atoms, and in order to suppress the dominant beam-related background
a pulsed beam is preferred. Monoenergetic electrons are searched for in the time interval between two 
pulses, when particles from the primary beam must be kept at a minimum. A figure of merit in this
kind of experiments is therefore the relative number of particles in the primary beam between two pulses,
called the extinction factor. 

In the $\mu \to 3e$ decay the three positrons, emitted simultaneously, share the muon rest energy, and
are emitted on a plane with the further constraint of zero total momentum. In analogy with the $\mu
\to e \gamma$ process the background from radiative decays is less important than the accidental 
coincidence of three particles from different muons, therefore a continuous positive muon beam is 
the preferred experimental solution.

\section{The MEG experiment}
The MEG experiment at PSI~\cite{MEG} aims at searching for the $\mu^+ \to e^+ \gamma$ decay with a
sensitivity of a few $\times 10^{-13}$. A beam of $3 \times 10^{7}$~$\mu^+$/sec is stopped on a thin
polyethylene target at the center of a superconducting magnet. The momentum and time of flight 
of positrons is measured by a set of drift chambers followed by a set of plastic scintillation counters,
while energy, conversion point and interaction time of $\gamma-$rays is measured by a single
volume liquid xenon detector, whose scintillation light is read by 846 UV-sensitive photo-multiplier
tubes.
An important feature of the MEG detector is its abundance of continuous and redundant calibration and 
monitoring which allow, for instance, a knowledge of the absolute energy scale of the photon 
detector with a systematics below $0.2 \%$.

MEG has being taking data since 2009 and the analysis of 2009--2011 data yields an upper limit on 
the branching ratio $\Gamma (\mu \to e \gamma ) / \Gamma( \mu \to e \nu \bar \nu) < 5.7 \times 
10^{-13}$ at 90\% confidence level~\cite{MEG2012}. This represents a factor of 5 improvement on the analysis 
of 2009--2010 data alone, due not only to the doubled statstics, but also to some hardware modifications 
(a new laser tracker for drift chamber alignment, a new BGO calibration detector) and analysis
improvements (better $\gamma-$ray pile-up unfolding on the photon side, and software noise reduction and 
revised track fitter for the positron side).   

The analysis of the 2012 data sample is in progress and futher statistics is being accumulated in 2013, 
when the experiment is expected to shut down due to the saturation of its sensitivity. In the meanwhile
the MEG collaboration has presented a proposal for an upgraded experiment~\cite{MEGUP} 
which aims at a sensitivity enhancement of one order of magnitude compared to the final MEG result.
The key features of the MEG upgrade are to increase the rate capability of all detectors (up to 
$7 \times 10^7$ $\mu^+$/sec) while improving 
the energy, angular and timing resolutions, for both the positron and photon arms of the detector. 
This is especially valid on the positron side, where a new low-mass, single volume, high granularity tracker 
is under development. A new highly segmented, fast timing counter array will replace the old system, 
to sustain the increased event rate.
\begin{figure}[htb]
\centering
\includegraphics[width=0.75\columnwidth]{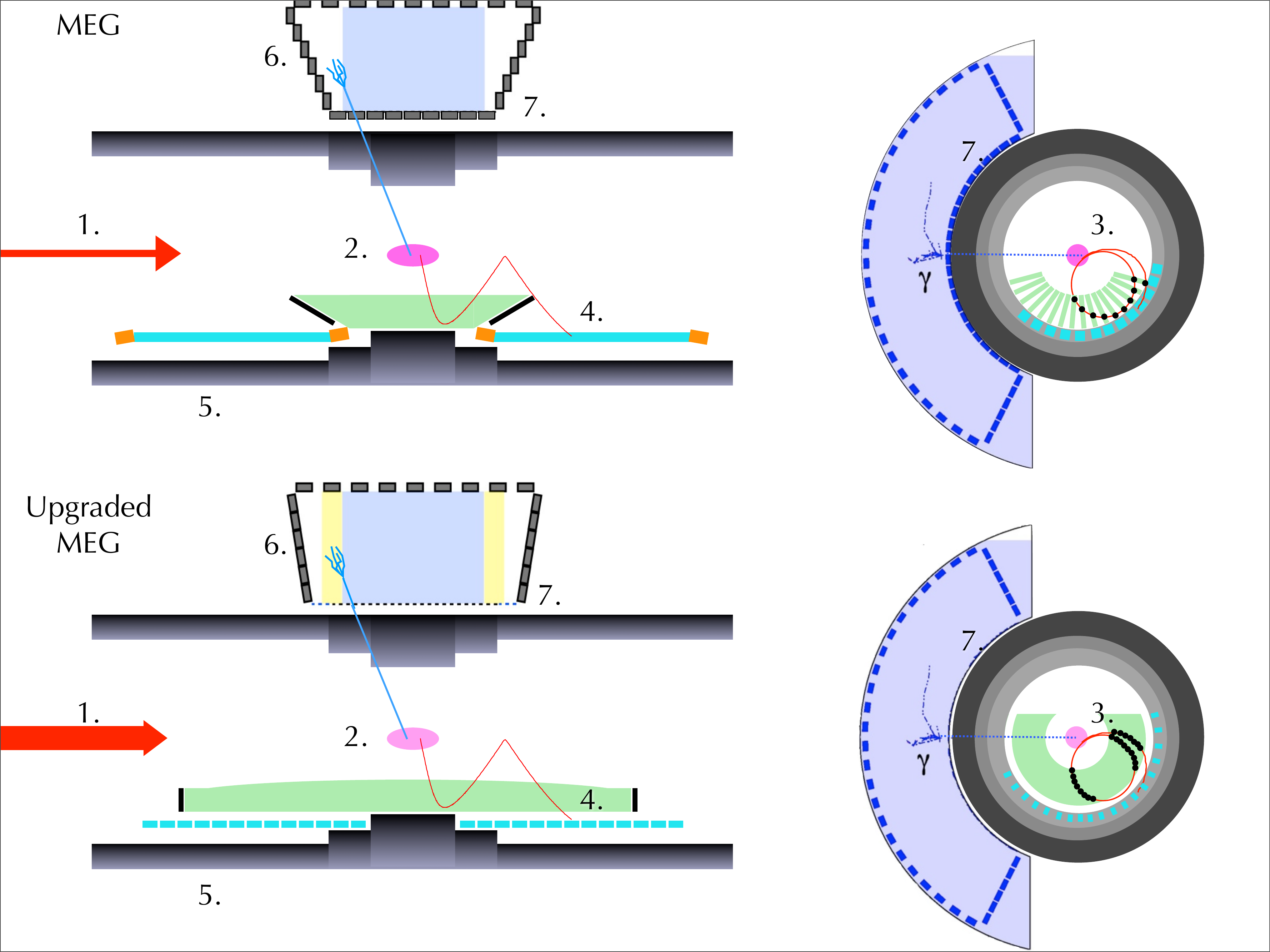}
\caption{A schematic representation of the MEG upgrade. An increased muon beam (1) impinges on a thinner
target (2). The new tracker (3) will provide more points and the coupling (4) with the timing counter will
be improved (5). The liquid xenon detector will be modified to provide more acceptance (6) and finer 
granularity (7).}
\label{fig:MEGupgrade}
\end{figure}

The photon-arm will also be improved by increasing the granularity of the liquid xenon detector
at the incident face, by replacing the current photomultiplier tubes (PMTs) with a larger number of 
smaller photosensors and optimizing the photosensor layout also on the lateral faces. Finally, a new 
DAQ scheme involving the implementation of a new combined readout board capable of integrating the various 
functions of digitization, trigger capability and splitter functionality into one condensed unit, 
will be necessary to cope with the increased rate and number of channels.

The upgraded detector is expected to be built in the years 2013-2015 and first engineering runs are 
planned at the end of 2015. A sensitivity of $5 \times 10^{-14}$ on the $\mu \to e \gamma$ branching
ratio is expected after data taking in the years 2016-2018. 
A schematic representation of the present and upgraded MEG detectors is 
shown in Figure~\ref{fig:MEGupgrade}.

\section{Mu3e at PSI}
\begin{figure}[htb]
\centering
\includegraphics[width=\columnwidth]{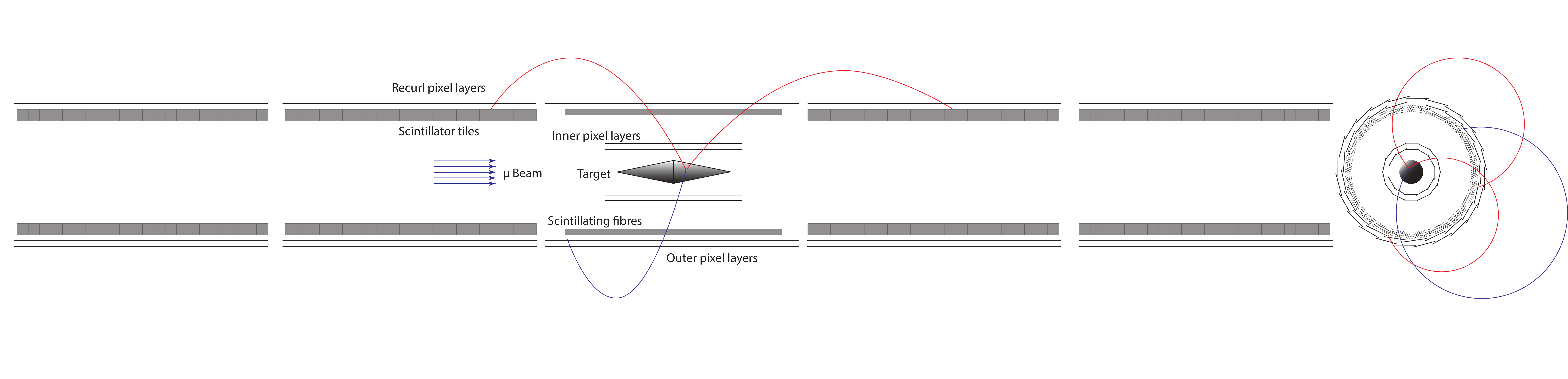}
\caption{A schematic of the complete Mu3e detector. Phase IA will have only the central barrel, while the
second and third pairs of stations will be added in phase IB and II respectively.}
\label{fig:mu3e}
\end{figure}

A proposal to search for the $\mu \to 3e$ decay was recently submitted at the PSI scientific committe%
~\cite{mu3e}.
A low energy muon beam is stopped on a double cone target. Positron trajectories are measured by a 
low material tracker made of ultra-thin silicon detectors based on high-voltage monolithic active pixel
sensors (HV-MAPS). A set of scintillating fibers and scintillating tiles will give a precise determination
of the electrons and positrons time of flight. Since all charged particles have very low momentum 
(from 15 to 50~MeV/c) the dominant effect that deteriorates the momentum resolution is multiple 
scattering. Minimizing the amount of material along the electrons/positrons trajectories is mandatory,
but a careful design of the tracking system (see Figure~\ref{fig:mu3e}) allows for partial cancelation
of multiple scattering effects at first order for most of the reconstructed trajectories.

The experimental quest is divided in phases: in phase IA only minimum detector configuration for early
commissioning, by means of a central silicon tracker only is envisaged. This will be supplemented 
in phase IB by scintillating recurl stations made of scintillating tiles and a fiber tracker. 
Adding the recurl stations will significantly enhance the momentum resolution and thus improve 
the suppression of background. The insertion of the fibre tracker and the tile detector stations gives 
a much better time resolution in comparison to the silicon pixel only.
The high time resolution will
allow running at the highest possible rate at the $\pi E 5$ muon beam line at PSI of 
$\approx 10^8~\mu^+$/sec. The sensitivity reach in this phase of the experiment of O(10$^{-15}$) 
will be limited by the available muon decay rate.

A phase II experiment is envisaged following the development of a new high intensity ($10^9~\mu^+$/sec)
beam line at PSI. The detector will be expanded with a second pair of recurl and tile station, 
which will allow the precise measurement of all particles crossing the inner silicon tracker.
The expected sensitivity on the $\mu \to 3e$ branching ratio is expected to reach the 
$10^{-16}$ level in this phase with a time scale which is competitive with that of the MEG upgrade.

\section{Muon to electron conversion experiments}
Aluminum is the candidate target for the research of coherent muon conversion on nuclei. A single
monoenergetic electron is present in the final state therefore the concept of accidental 
background is virtually absent in this case, and, unlike the $\mu \to e \gamma$ or $\mu \to 3e$ 
searches, there is no experimental wall at least until conversion rates of O($10^{-18}$). It is 
anticipated that muon-to-electron conversion will provide the ultimate sentitivity to CLFV detection.

The background comes from radiative muon decay in orbit or radiative muon capture (background neutron
and gamma rays will be produced), but mainly the beam related background (pion and electron contamination
in the primary beam) will be particularly harmful. To set the detector in a high-purity environment 
an idea of Dzhilkibaev and Lobashev~\cite{lobashev} of using a curved transport solenoid to bring 
muons and electrons from the production target to the detector together with a pulsed beam with challenging 
extinction is being pursued.

There are presently two big projects under development: the Mu2e project at the Fermi National Accelerator
Laboratory~\cite{Mu2e}, and the COMET project at J-PARC~\cite{comet}. 
\begin{figure}[htb]
\centering
\includegraphics[width=\columnwidth]{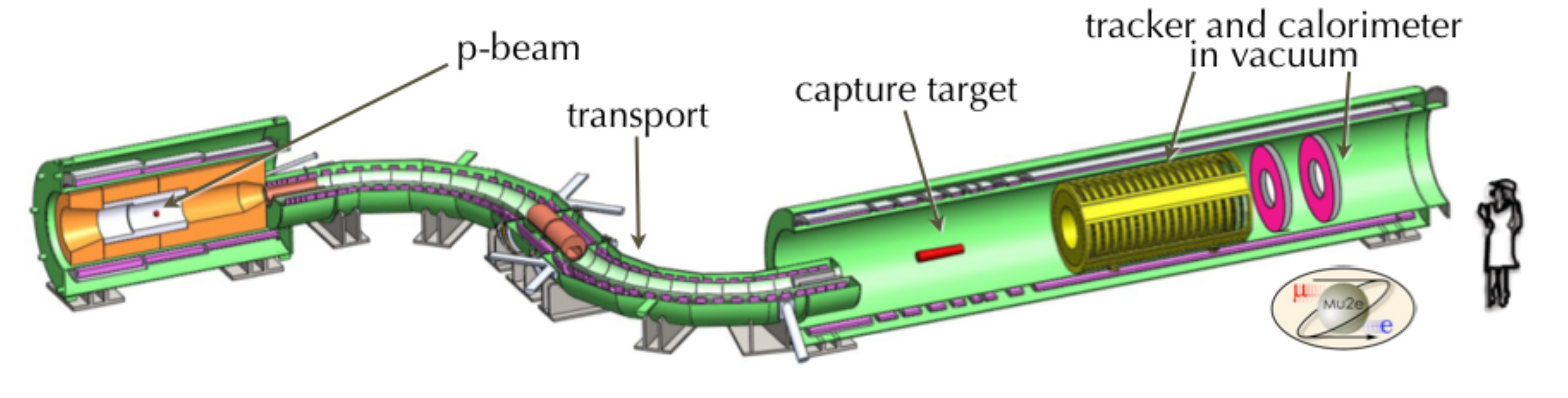}
\caption{A schematic of the Mu2e detector.}
\label{fig:mu2e}
\end{figure}
The two experiments are quite similar in the outline (see Figure~\ref{fig:mu2e}): 
a proton beam hits a target at the center
of a solenoid, where pions and muons are produced. Negative muons are transported along a curved
solenoid that acts as a very powerful momentum filter, therefore delivering a very clean low energy
muon beam onto the aluminum capture target placed downstream, in turn placed in a strong magnetic
field. Decay electrons are analyzed by a tracking device followed by a calorimetric/triggering
station. In the COMET detector, after the capture target, another curved solenoid is placed in
order to further decouple the electron spectrometer from the main beam transport line.

Both experiments are expected to reach a sensitivity of $3 \times 10^{-17}$ within 2020 and are 
heavily involved in detector development and extinction measurements.

COMET will follow a staged approach: COMET phase-I will be followed by COMET phase-II. For phase-I
the first 90 degrees of the muon beamline will be built with a twofold purpose: to make a direct
measurement of the proton beam extincion and other potential background sources for phase-II and to 
carry out a search for $\mu \to e$ conversion with a $3 \times 10^{-15}$ sensitivity.
After these measurements, the muon transport will be extended up to 180 degrees for the COMET Phase-II, 
and a $\mu \to e$ conversion search with a sensitivity of $3 \times 10^{-17}$ will be carried on with
an electron spectrometer and detectors.

In the meanwhile the DeeMe experiment at J-PARC~\cite{DeeMe}
 aims at making a $10^{-14}$ level measurement (a factor
of 100 better than the present limit) before 2017 by simplifying the beamline-detector geometry, 
having the same target for muon production and capture. A rotating silicon carbide is the candidate target 
material and will be followed by an electron spectrometer made of multi-wire proportional chambers.

\section{Summary}
The search for charged lepton flavor violation complements quark flavor physics measurements in search
for new or unexpected phenomena. Transitions involving muons are the most sensitive ones due to the
possibility of having very intense, low energy muon beams.
The MEG experiment recently improved the limit on the $\mu \to e \gamma$ decay down to $5.7 \times
10^{-13}$ at 90\% confidence level and it is presently finishing its data taking. An upgrade of 
the detector is in progress and will point to s ten times better sensitivity to be reached before 2018.

In parallel there are projects to search for the $\mu \to 3e$ decay in Europe, and to search for
the $\mu \to e $ conversion both in the USA and in Japan that will complement our knowledge in this 
sector within the next few years.

\Acknowledgements
I would like to thank the organizers of the FPCP2013 conference for their invitation. It was such a nice
experience to visit Brasil and especially Buzios. Discussions with colleagues from the MEG collaboration
are aknowledged.

\end{document}